# A Stable High Capacity Lithium-Ion Battery Using a Biomass-Derived Sulfur-Carbon Cathode and Lithiated Silicon Anode

Vittorio Marangon[a,†], Celia Hernández-Rentero[b,†], Dr. Mara Olivares-Marín[c], Prof. Vicente Gómez-Serrano[d], Prof. Álvaro Caballero[b], Prof. Julián Morales[b]*, Prof. Jusef Hassoun[a,e,f]*

[a] *Department of Chemical, Pharmaceutical and Agricultural Sciences, University of Ferrara, Via Fossato di Mortara 17, Ferrara 44121, Italy*

[b] *Department of Química Inorgánica e Ingeniería Química, Instituto de Química Fina y Nanoquímica, University of Córdoba, 14071 Córdoba, Spain*

[c] *Department of Ingeniería Mecánica, Energética y de los Materiales, University of Extremadura, Centro Universitario de Mérida, 06800 Mérida, Spain*

[d] *Department of Química Inorgánica, Facultad de Ciencias, University of Extremadura, 06006 Badajoz, Spain*

[e] *Graphene Labs, Istituto Italiano di Tecnologia, Via Morego 30 – 16163 Genova, Italy.*

[f] *National Interuniversity Consortium of Materials Science and Technology (INSTM), University of Ferrara Research Unit, University of Ferrara, Via Fossato di Mortara, 17, 44121, Ferrara, Italy.*

[†] Authors equally contributed.

*Corresponding Authors. E-mail addresses: iq1mopaj@uco.es (Julián Morales), jusef.hassoun@unife.it, jusef.hassoun@iit.it (Jusef Hassoun).

## Abstract

A full lithium-ion sulfur cell with a remarkable cycle life is achieved by combining environmentally a biomass-derived sulfur-carbon cathode and a pre-lithiated silicon oxide anode. X-ray diffraction

(XRD), Raman spectroscopy, energy dispersive spectroscopy (EDS), and thermogravimetry (TGA) of the cathode evidence the disordered nature of the carbon matrix in which sulfur is uniformly distributed with a weight content as high as 75 %, while scanning and transmission electron microscopy (SEM, TEM) reveal the micrometric morphology of the composite. The sulfur-carbon electrode exhibits in lithium half-cell a maximum capacity higher than 1200 mAh $g_S^{-1}$, reversible electrochemical process, limited electrode/electrolyte interphase resistance, and a rate capability up to C/2. The material shows a capacity decay of about 40% with respect to the steady state value over 100 cycles, likely due to the reaction with the lithium metal of polysulfides or impurities including P detected in the carbon precursor. Therefore, the replacement of the lithium metal with a less challenging anode is suggested, and the sulfur-carbon composite is subsequently investigated in the full lithium-ion sulfur battery employing a Li-alloying silicon oxide anode. The full-cell reveals an initial capacity as high as 1200 mAh $g_S^{-1}$, a retention increased to more than 79% for 100 galvanostatic cycles, and of 56 % over 500 cycles. The data reported herein well indicate the reliability of energy storage devices with extended cycle life employing high-energy, green and safe electrode materials.

**Table of Content**

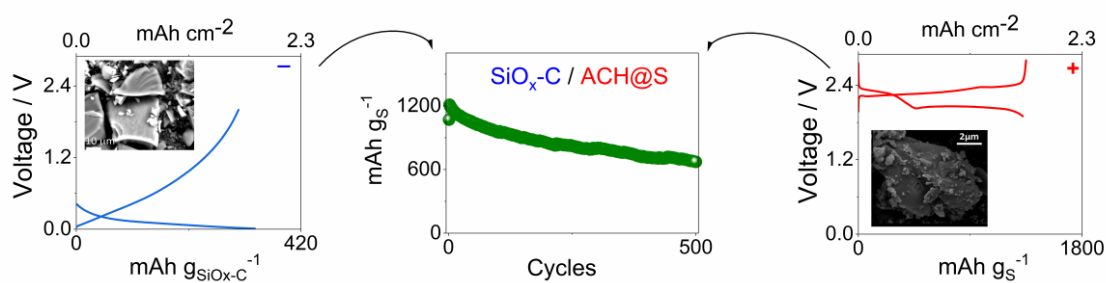

**A new Li-ion battery** exploiting a sulfur cathode derived from biomass and a silicon oxide alloying anode delivers a rather high capacity for an extended number of charge/discharge cycles

## Introduction

Lithium-sulfur (Li-S) battery is presently considered the most promising alternative for achieving higher energy and lower cost with respect to the world-wide employed lithium-ion battery.[1] The advantage in terms of energy density of a lithium cell using the sulfur cathode resides in the multi-electron electrochemical conversion process $16Li^+ + 16e^- + S_8 \rightleftarrows 8Li_2S$ which theoretically leads to 3730 Wh kg$^{-1}$ referring to sulfur mass.[2,3] Instead, the typical Li-ion cathode, e.g. layered metal oxide, (de)intercalates only a $x$ fraction of Li$^+$ ions ($0 < x < 1$) into the electrode structure during the electrochemical process and leads to a maximum theoretical energy density of about 900 Wh kg$^{-1}$ ($x = 0.8$) referring to the intercalated cathode mass.[4,5] The above mentioned electrochemical process of the Li-S battery actually involves the formation of lithium polysulfides with various chain length (Li$_2$S$_x$, $2 \leq x \leq 8$) as intermediates being high-order polysulfides able to dissolve into the electrolyte solution during cell discharge.[6] During subsequent charge, these mobile species can undergo a side reduction process at the lithium surface and subsequently migrate back to the cathode where they can be newly oxidized according to a continuous "shuttling" process leading to active material loss, electrodes degradation, decrease of delivered capacity, low coulombic efficiency, and, finally, to cell failure.[7,8] Among the various strategies adopted to limit side reactions at the lithium metal surface including the severe shuttle process of the Li$_2$S$_x$ species, the most relevant approach has proven that the addition of LiNO$_3$ as sacrificial agents to the electrolyte can protect the anode by forming a shielding solid electrolyte interphase (SEI) layer throughout a direct reduction reaction.[9–12] A further very promising approach has been represented by the entrapment of sulfur in carbon matrices of various natures and morphologies to directly limit the polysulfides dissolution.[13–17] Moreover, the employment of non-flammable and lowly-volatile electrolyte solutions such as end-capped glymes[18–20], ionic liquids[21,22] or polymers[23] has been indicated to provide a safe and stable cycling behavior. Despite the recent notable improvements of the Li-S technology,[24] the use of a lithium-metal anode may still represent a potential safety issue that could prevent the actual use of these promising high-energy devices. On the other hand, during eighties the reliability of the Li-ion batteries has been successfully achieved by replacement of the energetic lithium-metal anode (3860 mAh g$^{-1}$, -3.04 V *vs.* SHE) with graphite to avoid the growth of metallic dendrites promoted by a heterogeneous metal deposition upon charge, possibly leading to short-circuits and consequent cell failure.[25] These researches have promoted the commercialization of the lithium-ion battery, and have been awarded in 2019 by the Nobel Prize in Chemistry.[26,27] Thus, the application of the Li-ion concept through the replacement of the metallic lithium with a stable and non-reactive anode based on lithium intercalation,[28] conversion,[29] or alloying[30] may actually represent an attractive compromise to safely exploit the multi-electron conversion process of the Li-S battery.[31–36] In particular, Li-alloy with Sn[37,38] and Si[39] or their oxides[40–44] exploiting the nanostructured morphology have revealed higher capacity compared to graphite (372 mAh g$^{-1}$), with values ranging from 500 to 1000 mAh g$^{-1}$, due to the multiple lithium ion exchange per molar unit of metal. Another raising point has been represented by the sustainability of the new energy storage devices due to climate change issues during the recent years which focused the attention on the necessity of eco-friendly materials.[45,46] In this respect, outstanding studies have demonstrated that carbon-based electrodes obtained from the recycle of bio-waste products may represent suitable alternative to enable sustainable and, at the same time, high-performance energy storage devices.[47–49] Indeed, Li-S batteries relying on cathode materials derived from crab shells,[50] peanut shells,[51] olive stones,[52] rice husks,[53] cherry pits,[54] bamboo,[55] brewing waste,[56] and even jellyfish umbrellas have been proposed as possible alternatives.[57] Taking in mind these achievements, we have explored herein the concept of a full lithium-ion sulfur battery based on sustainable materials according to the most recent worldwide plans of the *green economy*, in particular within the UE community.[58] Hence, a pre-lithiated silicon oxide-based anode characterized by well suitable cell performances and a biomass-derived sulfur-carbon cathode have been coupled in a new long life ecofriendly energy storage device.[59,60] The new sulfur-carbon composite is initially investigated in terms of structure, morphology, thermal behavior and applicability in lithium half-cell. Furthermore, direct lithiation process of the silicon oxide-based anode is exploited to allow the combination of the two electrodes, leading to environmentally compatible lithium-ion sulfur battery that may actually allow the optimization of green and high-performance energy storage system alternative to the conventional lithium-ion battery.

**Results and discussion**

The structural features of the carbon-sulfur composite (AC-H@S) are initially investigated by means of XRD and the results are reported in Figure 1a. The AC-H@S pattern exhibits the sulfur ($S_8$) signals between $2\theta = 20$ and $60°$ and without any crystallographic evidence corresponding to graphite which is usually observed at about $2\theta = 26°$, as expected by the disordered nature of the activated carbon precursor (AC-H).[60,61] Furthermore, the exclusive presence of the sulfur signals implies the absence of impurities and, thus, the effectiveness of the synthesis pathway. The disordered nature of the carbonaceous frame of AC-H@S is confirmed by Raman spectroscopy reported in Figure 1b. Indeed, the presence of broad D (~1350 cm$^{-1}$) and G (~1600 cm$^{-1}$) bands, the related intensity ratios ($I_D/I_G$) of 0.95, as well as the absence of a defined 2D band generally observed around 2700 cm$^{-1}$ for graphitic structures,[62,63] indicate a large ratio of structural defects.[60,64] The Raman spectrum also identifies the sulfur hosted in the composite, which is represented by the narrow peak centered at about 473 cm$^{-1}$.[65] The actual amount of sulfur in AC-H@S is detected through TGA in Figure 1c, which reveals a sulfur content as high as 75 % that is expected to enable high energy density of lithium-metal and lithium-ion cells.[66] Furthermore, the DTG reported in the bottom panel of Fig. 1c evidences that the sulfur weight loss evolves through two subsequent steps, among which the first and major one is centered at 300 °C (51 % of the total sulfur loss) while the second one extends between 300 and 400 °C (24 % of the total sulfur loss). The first step is likely ascribable to sulfur located on the external carbon surface, instead the second one can be related to the active material hosted within the micro-porous carbonaceous structure.[60,67,68] It is worth mentioning that the above electrode architecture may actually enhance the electrical contact between the active material (i.e., sulfur) and the conductive matrix (the carbon frame), thus shortening the electron pathway and enabling the kinetics of the lithium-sulfur electrochemical conversion process and the cell cycling.[69] The electron microscopy of the AC-H@S reported in Fig. 1 shows a sample formed by submicron flakes (TEM in Fig. 1d) aggregated into particles with a size ranging from 1 μm or smaller to about 10 μm (SEM in Fig. 1e). Furthermore, the EDS analyses carried out on the TEM image of Fig. 1f display a uniform elemental distribution of sulfur (Fig. 1g) and carbon (Fig. 1h) forming the electrode matrix, as well as the above mentioned traces of phosphorous (Fig. 1i) due to the $H_3PO_4$ activating agent used for the AC-H precursor synthesis.[60,70] The observed morphology, characterized by the concomitant presence of sulfur-carbon particles with a wide size range, may actually play an important role in achieving enhanced battery performances of the AC-H@S composite, since the small sulfur particles generally enable a high capacity values while the large ones can act as an active material reservoirs and enable stable cycling.[71]

**Figure 1**

The electrochemical features of the AC-H@S electrode are investigated in lithium half-cell by cyclic voltammetry (CV) and electrochemical impedance spectroscopy (EIS), as depicted in Figure 2. The first CV profile (Fig. 2a) shows the typical signature of the Li-S conversion process detected by two discharge peaks at 2.25 and 2.0 V *vs.* $Li^+/Li$, corresponding to the formation long-chain lithium polysulfides ($Li_2S_8$ and $Li_2S_6$) and short-chain ones ($Li_2S_x$, $2 \leq x \leq 4$), respectively, reversed in a broad double-peak between 2.3 and 2.5 V *vs.* $Li^+/Li$ during charge which indicates the conversion of the polysulfides back to lithium and sulfur.[72] The subsequent voltammetry cycles reveal a shift of the discharge peaks to higher potential values, suggesting the occurrence of an activation process that leads to a lower polarization between charge and discharge.[69] The activation process is often observed taking place in the first cycles of lithium-sulfur batteries and is usually associated to rearrangements of the sulfur electrode accompanied by a structural reorganization generally leading to the stabilization of electrode/electrolyte interphase and the enhancement of the electrode conductivity.[69] Furthermore, the narrow discharge/charge signals exhibited by AC-H@S, as well as the notable overlapping of the potential profiles, suggest an efficient conversion process characterized by fast kinetics. Additional details on the behavior of the AC-H@S electrode in lithium half-cell are provided by the EIS measurements reported in Fig. 2b carried out upon CV. The Nyquist plots are analyzed by non-linear least squares (NLLS) method to obtain the corresponding equivalent circuit formed by resistive (R) and constant phase elements (CPE, Q), and identified by the $R_e(R_iQ_i)Q_w$ model as reported in Table 1.[73,74] In detail, $R_e$ is the electrolyte resistance, identified by the high frequency intercept in the Nyquist plots, $R_i$ and $Q_i$ parallel elements $(R_iQ_i)$ represent the single or multiple high-medium frequency semicircles and account for the electrode/electrolyte interphase, while $Q_w$ indicates the Warburg-type $Li^+$ ions diffusion which is observed as a tilted line at low frequency values in the Nyquist plot.[66] The results of NLLS analyses reported in Table 1 reveal that the above mentioned favorable activation process of the AC-H@S electrode upon the first CV cycle is well supported by the decrease of total interphase resistance ($R_{tot}$, given by the sum of the $R_i$ elements) as well as by the corresponding Nyquist plot shrinks. Indeed, the cell exhibits a total resistance of about 40 Ω at the OCV (inset in Fig. 2b), and a drop down to around 8 Ω after 1 cycle and 6 Ω after 10 cycles (Fig. 2b). Further modifications of the electrode/electrolyte interphase upon CV may be inferred by the change of the Nyquist plot shape and the increase of the $(R_iQ_i)$ elements number in the corresponding equivalent circuit (Table 1), which is likely ascribed to a change of the electrode morphology.[69]

**Table 1**

**Figure 2**

The electrochemical performances of the AC-H@S electrode in lithium half-cell are evaluated in Figure 3 through galvanostatic tests at increasing current from C/10 (1C = 1675 mA g$^{-1}$) to C/8, C/5, C/3, C/2, 1C and 2C (Fig. 3a and 3b), and at the constant rate of C/3 for 100 cycles (Fig. 3c and 3d). The evolution of the voltage profiles reported in Fig. 3a evidences that the Li/AC-H@S cell at a C-rate lower than 1C operates according to the CV of Fig. 2 with two discharge plateaus centered at 2.3 and 2.0 V due to the reduction of sulfur to lithium polysulfides, and merging plateaus between 2.3 and 2.4 V due to the subsequent oxidation. The cell reveals the excepted increase of the discharge/charge polarization by raising the current from C/10 to C/2, however a further increase to 1C and 2C turns into the deactivation of the electrochemical process due to excessive overvoltage which is indicated by the concomitant drop of the delivered capacity. Nonetheless, the cycling trend depicted in Fig. 3b shows for the AC-H@S electrode stable capacity values of 1200, 1180, 1100 and 1000 mAh $g_S^{-1}$ at C/10, C/8, C/5 and C/3, respectively, and between 890 and 780 mAh $g_S^{-1}$ at C/2. After the abrupt decay of the delivered capacity below 300 and 150 mAh $g_S^{-1}$ at 1C and 2C, respectively, the cell recovers about 92 % of the initial value when the C-rate is lowered back to C/10, thus suggesting a good stability of the active material by changing currents. Fig 3c shows selected voltage profiles related to the Li/AC-H@S cell characterized at the constant current of C/3 for 100 cycles. Interestingly, the poor capacity exhibited during the first cycle (400 mAh $g_S^{-1}$), as well as the anomalous evolution of the corresponding discharge/charge plateaus, may be attributed to an initial low conductivity of the electrode/electrolyte interphase, which is improved by the activation process after 1 cycle as already discussed in Fig. 2. On the other hand, the electrode exhibits a remarkable capacity, with starting values higher than 1200 mAh $g_S^{-1}$, and a coulombic efficiency approaching 99 % at the steady state (Fig. 3d). However, the half-cell shows a retention limited to 60 % upon 100 discharge/charge cycles, which may be ascribed to the reaction of the lithium metal with polysulfides or impurities such as phosphorous dissolved from the activated carbon matrix (see EDS and related discussion in Fig. 1). Therefore, we may assume that the AC-H@S electrode has suitable performance for battery application, in particular for Li-ion cell in which the above issues ascribed to the presence and reactivity of the lithium metal can be actually mitigated. The AC-H@S electrode is subsequently coupled with a silicon oxide-based anode ($SiO_x$-C) in a full lithium-ion sulfur battery. Prior to using, the $SiO_x$-C electrode was activated by galvanostatic cycling in lithium cell (see experimental section) in order to obtain the lithiated (charged) $Li_ySiO_x$-C anode which is a suitable $Li^+$ ions reservoir in the full Li-ion sulfur cell.[75] The voltage profiles of the galvanostatic test performed on the $Li_ySiO_x$-

C/AC-H@S full-cell at the constant current rate of C/5 (1C = 1675 mA g$_s^{-1}$) reported in Fig. 3e reveal an electrochemical process centered at about 1.8 V. The discharge and charge processes evolve according to the combination between the typical voltage shapes associated to the multi-step conversion process of the Li-S battery[3] and the (de-)alloying mechanism of the Li/SiO$_x$-C cell.[59] During the first cycle the discharge exhibits two broad plateaus extending in voltage intervals of 1.9 – 2.3 V and 1.1 – 1.6 V, which are reversed into a sloping charge profile evolving between 1.5 and 2.35 V. Interestingly, the subsequent cycles exhibit the gradual fragmentation of the charge plateau into 3 different processes taking place at 1.7, 2.2 and 2.45 V. This trend is likely ascribed structure rearrangements and consolidation of a stable interphase at the surface of both electrodes, as suggested by CV and EIS measurements in Fig. 2 and by previous works.[76] This complex process is likely reflected into an increase of delivered capacity upon the first cycle, as also evidenced by the cycling trend shown in Fig. 3f. Indeed, the Li$_y$SiO$_x$-C/AC-H@S cell exhibits 1070 mAh g$_s^{-1}$ during the first cycle that raise up to 1210 mAh g$_s^{-1}$ during the second one. The Li-ion sulfur cell shows a relevant capacity retention, in particular if compared to the corresponding Li-sulfur half-cell. Hence, the corresponding plots in Fig. 3f (Li-ion cell) and Fig. 3d (Li-S cell), indicate a retention over 100 cycles of about 79% for the former and of 60% for the latter. Furthermore, the test extended for 500 cycles in the Li-ion cell (Fig. 3f) reveals a residual capacity as high as 670 mAh g$_s^{-1}$ leading to a retention of 56 % of the maximum value, and a coulombic efficiency higher than 92%. Therefore, considering an average operating voltage of about 1.8 V and a capacity of 670 mAh g$_s^{-1}$ after 500 cycles, we can estimate that the Li$_y$SiO$_x$-C/AC-H@S cell can still hold upon this challenging test a theoretical specific energy density of about 1200 Wh kg$_s^{-1}$, which could lead to practical value of 400 Wh kg$^{-1}$ by taking in consideration a correction factor of 1/3 that includes all the inactive components of the cell.[77] This stable cycling behavior, the remarkable delivered capacity and energy, and the proper voltage evolution of the cell are herein achieved by tuning the negative-to-positive (N/P) ratio with a very limited anode excess, that is, 1.04 (See Figure S1 in Supporting Information). The N/P ratio approaching the unity may in fact favor the achievement of optimized full-cell performances as indicated in previous papers.[4,32]

**Figure 3**

With the aim of further understanding the behavior of the full Li-ion sulfur cell, we investigate hereafter the morphological and structural features of the AC-H@S and $SiO_x$-C electrodes at the pristine state and after cycling. Indeed, an additional full AC-H@S/$Li_ySiO_x$-C cell is assembled with an anode achieved by chemical pre-lithiation of the $SiO_x$-C material through direct contacting the electrode disk with a lithium foil soaked with the electrolyte to reach the $Li_ySiO_x$-C alloy (see experimental section for further details). This activation pathway allows a rapid and efficient lithiation of the electrode,[78] as demonstrated by Figure 4. The selected voltage profiles of lithium half-cells assembled either with a pristine $SiO_x$-C electrode (Fig. 4a) or with $Li_ySiO_x$-C electrodes achieved by chemical pre-lithiation at various time regimes, that is, 30 minutes (Fig. 4b), and 1, 2 and 14 hours (Fig. 4c, d and e respectively) show that the capacity delivered at the first discharge of the half-cells, corresponding to the charge step in full-cell, decreases by increasing the contact time, thus indicating the progressive lithiation "activation" of the $SiO_x$-C electrode (see histogram in Fig. 4f). The $Li_ySiO_x$-C/AC-H@S full-cell is then assembled by coupling a fresh cathode with an anode chemically activated for 48 hours to ensure the complete lithiation, and galvanostatically cycled at the constant current rate of C/5 for 20 cycles (Fig. 4g).

**Figure 4**

Subsequently, the two electrodes are recovered after disassembling the cycled cell and characterized along with pristine AC-H@S and $SiO_x$-C disks by SEM and XRD as displayed in Figure 5, while the corresponding elemental distribution is detected by EDS (Figures S2 and S3, respectively, in the Supporting Information). The SEM image related to the pristine AC-H@S cathode (Fig. 5a) reveals the presence of micrometric sulfur (bright domains) and $H_3PO_4$-activated carbon particles (grey domains) uniformly distributed on the electrode surface (Fig. S2a), as evidenced by the EDS elemental maps of sulfur (Fig. S2b) and phosphorous (Fig. S2c), while the one related to fluorine (Fig. S2d) evidences the PVDF added to the electrode formulation as the polymer binder (see experimental section). On the other hand, the pristine $SiO_x$-C shows large particles with size exceeding 20 μm (Fig. 5b) formed by a carbon matrix containing silicon oxide particles, as identified by the EDS elemental maps of C (Fig. S3a), Si (Fig. S3b) and O (Fig. S3c).[59] Substantial modifications of the morphology can be observed in Fig. 5c and 5d, which display the SEM images of the AC-H@S and $SiO_x$-C electrodes, respectively, after cycling. Indeed, AC-H@S shows a surface apparently filled by active material or carbon particles different than the pristine one (compare Fig. 5a and Fig 5c), while $SiO_x$-C displays particles having similar shape with respect to the pristine state however with smaller size (compare Fig. 5b and Fig 5d), as likely associated with the unavoidable volume changes due to the Li-Si (de-)alloying process and possible partial fragmentation. Both AC-H@S and $SiO_x$-C exhibit the presence of a bright uniform surface layer suggesting the growth of a SEI formed by carbon (EDS Figs. S2e and S3e), sulfur (Fig. S2f and inset in Fig. S3h), fluorine (Figs. S2h and S3h) and oxygen (inset in Fig. S2h and Fig. S3g). The above elemental composition of the SEI layer can be attributed to the decomposition of the DOL and DME ether chains or the LiTFSI conductive salt in the electrolyte,[79,80] to the electrodeposition of amorphous sulfur upon charge on the AC-H@S surface,[66] as well as to possible side reaction of lithium polysulfides with the lithiated $SiO_x$-C electrode.[81] Interestingly, the XRD pattern related to the AC-H@S electrode exhibits at the pristine state the broad peak centered at 2θ = 26 ° ascribed to the porous carbon-cloth electrode (see experimental section) and the typical crystalline sulfur signals between 2θ = 20 and 60 ° which disappear in the pattern of the cycled electrode (Fig. 5e). Furthermore, both the patterns related to the $SiO_x$-C electrode (before and after cycling) show exclusive peaks ascribed to the copper support in addition to sulfur impurity detected by EDS (inset in Fig. S3h), as well as the unaltered amorphous structure of the carbon-embedded silicon oxide particles (Fig. 5f). Therefore, the SEM/EDS analyses and the XRD measurements suggest specific morphological modifications and structural stability of the two electrodes upon cycling in full-cell, and the formation of a protecting SEI layer through a series of favorable side reactions at the electrode/electrolyte interphase.

**Figure 5**

**Conclusions**

A carbon-sulfur composite indicated as AC-H@S has been synthesized using a bio-residues carbonaceous precursor and characterized in view of possible application in a sustainable Li-ion sulfur battery. XRD measurement performed on the composite powder evidenced the absence of excessive impurities and the predominant presence of sulfur, as well as the disordered nature of the carbon frame as confirmed by Raman spectroscopy. TGA of the AC-H@S detected a sulfur content as high as 75 % which is allowed by the micro-porous structure of the activated carbon. CV and EIS tests performed on lithium half-cell suggested fast kinetics of the electrochemical conversion process and a remarkable conductivity of the electrode/electrolyte interphase upon activation. In particular, CV profiles identified two reversible and narrow discharge peaks at 2.25 and 2.0 V *vs.* $Li^+/Li$ reversed into a charge process extending from 2.3 to 2.5 V *vs.* $Li^+/Li$, with interphase resistance values measured through EIS of 40 Ω at the OCV decreasing down to 6 Ω upon 10 CV cycles. Galvanostatic cycling tests carried out on lithium half-cells have shown maximum capacity values exceeding 1200 mAh $g_s^{-1}$, coulombic efficiency approaching 99 %, and a rate capability extending up to C/2. Despite the suitability of the AC-H@S for battery application, the half-cell suffered by an excessive decay of the capacity by cycling due to side reactivity of the lithium metal with polysulfides and phosphorous impurities possibly dissolved into the electrolyte upon cathode operation. Therefore, the lithium electrode was replaced by a Li-alloy anode based on silicon oxide into amorphous carbon to achieve the $Li_ySiO_x$-C/AC-H@S cell exploiting the Li-ion configuration. The new battery revealed at C/5 a sloped voltage signature centered at about 1.8 V in line with the combination of the multi-step sulfur conversion and the (de-)alloying process of the $Li_ySiO_x$-C anode. Furthermore, the $Li_ySiO_x$-C/AC-H@S full-cell delivered a maximum capacity of about 1200 mAh $g_s^{-1}$ retained slightly below 60 % over 500 cycles, with a final theoretical energy density of about 1200 Wh $kg_S^{-1}$ and an estimated practical value of 400 Wh $kg^{-1}$. Such a notable performance has been herein achieved by using enhanced anode and cathode materials, and by properly tuning their negative-to-positive (N/P) ratio to a value approaching the unity (i.e., of about 1.04), thus concomitantly allowing long cycle life and high delivered capacity. Accordingly, the structural retention and the formation of a suitable SEI layer have been actually observed by performing *ex*-situ XRD and SEM-EDS measurements on pristine electrodes and on materials recovered from the $Li_ySiO_x$-C/AC-H@S cell upon cycling. Therefore, this study may represent a step forward to achieve an alternative Li-ion battery employing environmentally friendly materials, such as sulfur, bio-waste derivatives and silicon, characterized by high-energy and extended cycle life.

**Experimental section**

*Synthesis of the carbon precursor*

The carbon precursor exploited herein was obtained by treatment of bio-mass residues of cherry pits provided by Asociación de Cooperativas del Valle del Jerte (Cáceres province, Spain), as reported in a previous work.[60] Accordingly, the treated cherry pits powder was activated through phosphoric acid ($H_3PO_4$) treatment of the precursor,[70] and the obtained sample was subsequently annealed under $N_2$ and indicated as AC-H. The detailed chemical-physical characterization of the AC-H carbon precursor is reported elsewhere.[60]

*Synthesis of the carbon-sulfur composite*

The carbon-sulfur composite was obtained by infiltrating sulfur in the AC-H carbon precursor via *in situ* disproportionation of sodium thiosulfate pentahydrate ($Na_2S_2O_3 \cdot 5H_2O$) in acidified aqueous solution. Accordingly, 3 g of $Na_2S_2O_3 \cdot 5H_2O$ (Sigma-Aldrich) were dissolved in a solution composed of 150 mL of $H_2O$ and 4.5 mL of a Triton X-100 solution (1 vol%), that is, a polymer surfactant that avoids sulfur agglomerates and allows controlled sulfur particles size. Separately, 100 mg of carbon sample (AC-H) were dispersed in 100 mL of $H_2O$ and sonicated for 1 hour. Subsequently, the two solutions were mixed together and heated at 70 °C with the aid of a silicon oil bath, and 15 mL of HCl (12M) were slowly added under vigorous magnetic stirring to achieve the following reaction (1):

$$Na_2S_2O_3 \cdot 5H_2O \,(aq) + 2HCl\,(aq) \rightarrow 2NaCl\,(aq) + SO_2\,(g) + S\,(s) + 6H_2O \qquad (1)$$

After 15 minutes, the silicon oil bath was removed and the mixture was left under mild magnetic stirring at room temperature for 24 hours. Finally, the obtained carbon-sulfur composite was washed repeatedly with $H_2O$, ethanol and acetone via centrifugation to remove HCl and Triton X-100, and then dried at 50 °C in an oven overnight. The final carbon-sulfur sample is then indicated as AC-H@S.

*Carbon-sulfur powder characterization*

The structural features of the AC-H@S powder were investigated through X-ray diffraction (XRD) and Raman spectroscopy. XRD pattern was obtained by a Bruker D8 Discover X-ray diffractometer exploiting monochromatic Cu Kα radiation to scan the 2θ range between 10° and 80° by using a step size of 0.04° and a rate of 1.05 s step$^{-1}$.

Raman spectroscopy was carried out under ambient conditions by using a Renishaw inVia Microscope equipped with a Renishaw CCD Camera (578 × 400) detector and a 532 nm edge in line focus mode laser.

The sulfur content of AC-H@S was determined by thermogravimetric analysis (TGA) performed through a Mettler Toledo TGA/DSC-1 from 30 to 800 °C with a heating rate of 10 °C min$^{-1}$ under $N_2$ flow.

Sample morphology was studied by scanning electron microscopy (SEM) and transmission electron microscopy (TEM) using a JEOL JSM-7800F and a JEOL 2010 electron microscope operating at 200 kV equipped with an Orius Gatan CCD camera, respectively. The elements distribution of the AC-H@S composite was evaluated via energy-dispersive X-ray spectroscopy (EDS), which was performed on the TEM images through a X-ACT Cambridge Instrument analyzer.

*Synthesis of the SiO$_x$-C material*

The synthesis of the SiO$_x$-C composite was achieved through sol-gel method, as reported elsewhere.[59] 18 g of resorcinol were mixed with 58.5 g of formaldehyde at room temperature until a homogeneous mixture was obtained. Subsequently, 21 g of tetraethyl orthosilicate (TEOS) were added to the solution, which was then heated at 70 °C. The dropwise addition of 2 mL of HCl (1M) to the heated solution catalyzed the formation of a semitrasparent pink gel, which was aged for 24 hours at rom temperature and then cut into pieces, washed with ethanol to remove residual HCl, and finally annealed at 1000 °C for 10 hours under Ar-H$_2$ (5%) flow. The black powder obtained was ground in a mortar.

*Electrodes preparation and electrochemical characterization*

The electrodes slurries were prepared by mixing 80 wt% of the active material, either AC-H@S or SiO$_x$-C, 10 wt% of Super P carbon (Timcal) as conductive agent, and 10 wt% of polyvinylidene fluoride (Solef ® 6020 PVDF) as polymer binder, dispersed in 1-methyl-2-pyrrolidone (NMP, Sigma-Aldrich). The slurries containing AC-H@S and SiO$_x$-C were coated on a carbon cloth foil (GDL, ELAT LT1400W, MTI Corp.) and a Cu foil (MTI Corp.), respectively, by using a doctor blade (MTI Corp.). The GDL carbon cloth was used for the sulfur electrode to achieve optimal performances due to its better textural properties compared to common Al [82,83]. Then, the electrode films were heated at 50 ºC for 5 h under air and subsequently cut into disks of 14 mm diameter, which were dried at 45 ºC overnight under vacuum to remove residual traces of water and NMP. The active material loading on the final electrodes was of about 1.3 mg cm$^{-2}$ for AC-H@S and 5.3 mg cm$^{-2}$ for SiO$_x$-C.

The electrochemical processes of the AC-H@S and SiO$_x$-C composites were analyzed in 2032 coin-type cells (MTI Corp.) assembled in an Ar-filled glove box (MBraun, O$_2$ and H$_2$O content below 1 ppm) by stacking either an AC-H@S or a SiO$_x$-C disk as the positive electrode, a 16 mm diameter Celgard foil soaked with the electrolyte as the separator, and a 14 mm diameter lithium metal disk as the negative electrode. The electrolyte solution exploited in this work was obtained by solvating lithium bis(trifluoromethanesulfonyl)imide (LiTFSI, LiN(SO$_2$)$_2$(CF$_3$)$_2$, 99.95% trace metals basis, Sigma-Aldrich) conductive salt and lithium nitrate (LiNO$_3$, 99.99% trace metals basis, Sigma-Aldrich) sacrificial agent in a solution of 1,3-dioxolane (DOL, C$_3$H$_6$O$_2$, anhydrous, contains ca. 75 ppm BHT as inhibitor, 99.8%, Sigma-Aldrich) and 1,2-dimethoxyethane (DME, CH$_3$OCH$_2$CH$_2$OCH$_3$, anhydrous, 99.5%, inhibitor-free, Sigma-Aldrich) mixed in a 1:1 w/w ratio. LiTFSI and LiNO$_3$ were added to the DOL:DME 1:1 w/w solution to obtain a final concentration of 1 mol kg$^{-1}$ for each salt, as referred to the solvent mass. Prior to electrolyte preparation, LiTFSI and LiNO$_3$ were dried under vacuum to 110 and 80 °C, respectively, for 3 days to remove any trace of water, while DOL and DME were dried with the aid of molecular sieves (3 Å, rod, size 1/16 in., Honeywell Fluka) until a water content below 10 ppm was obtained, as measured by a Karl Fischer 899 Coulometer (Metrohm).

Cyclic voltammetry (CV) and electrochemical impedance spectroscopy (EIS) tests were performed on a Li/DOL:DME 1:1 w/w, 1 mol kg$^{-1}$ LiTFSI, 1 mol kg$^{-1}$ LiNO$_3$/AC-H@S cell through a VersaSTAT MC Princeton Applied Research (PAR) analyzer. CV measurements were carried out in the 1.8 – 2.8 V *vs.* Li$^+$/Li potential range by using a scan rate of 0.05 mV s$^{-1}$, while EIS measurements were performed at the open circuit (OCV) condition and after 1, 5 and 10 CV cycles in the 500 kHz – 100 mHz frequency range by using a 10 mV alternate voltage signal amplitude. The recorded impedance spectra were analyzed by non-linear least squares (NLLS) method through a Boukamp tool and only fitting with a $\chi^2$ value of the order of 10$^{-4}$ or lower were considered suitable.[73,74]

Galvanostatic cycling tests were performed by using a MACCOR series 4000 battery test system. The electrochemical performances of Li/DOL:DME 1:1 w/w, 1 mol kg$^{-1}$ LiTFSI, 1 mol kg$^{-1}$ LiNO$_3$/AC-H@S half-cells were evaluated through prolonged discharge/charge cycling at the constant current rate of C/3, and through rate capability measurements employing current values of C/10, C/8, C/5, C/3, C/2, 1C and 2C by increasing the current rate every 5 cycles and lowering it to the initial value of C/10 after 35 cycles (1C = 1675 mA g$_S^{-1}$). The 1.9 – 2.8 V voltage range was employed from C/10 to C/2 rates, while tests at 1C and 2C were carried out between 1.8 and 2.8 V. Both specific current and specific capacity were referred to the sulfur mass.

The lithium-ion sulfur cell was assembled by coupling the AC-H@S electrode as cathode with a pre-lithiated $SiO_x$-C anode ($Li_ySiO_x$-C) in 2032 coin-type cells (MTI Corp.) using the $Li_ySiO_x$-C/DOL:DME 1:1 w/w, 1 mol kg$^{-1}$ LiTFSI, 1 mol kg$^{-1}$ LiNO$_3$/AC-H@S configuration. $SiO_x$-C electrodes were pre-activated through 30 discharge-charge cycles by employing a constant current rate of 50 mA g$^{-1}$ in the 0.01–2.0 V voltage range in Li/DOL:DME 1:1 w/w, 1 mol kg$^{-1}$ LiTFSI, 1 mol kg$^{-1}$ LiNO$_3$/$SiO_x$-C cells. The $Li_ySiO_x$-C electrodes were recovered from the above cell disassembled at 0.01 V, washed by using DME and dried under vacuum for 30 minutes. Galvanostatic cycling tests were performed on the $Li_ySiO_x$-C/DOL:DME 1:1 w/w, 1 mol kg$^{-1}$ LiTFSI, 1 mol kg$^{-1}$ LiNO$_3$/AC-H@S full-cell within the 0.1–2.8 V voltage window at a current rate of C/5 (1C = 1675 mA g$_s^{-1}$). Both specific current and specific capacity of the full-cell were referred to sulfur mass.

Furthermore, chemical lithiation of the $SiO_x$-C electrode was performed to achieve a suitable condition for practical application of the $Li_ySiO_x$-C material in full-cell. The above lithiated anode was achieved by direct contacting the $SiO_x$-C electrode with a lithium foil soaked with the DOL:DME 1:1 w/w, 1 mol kg$^{-1}$ LiTFSI, 1 mol kg$^{-1}$ LiNO$_3$ electrolyte under a pressure of 2 kg cm$^{-2}$ for selected time regimes.[78] The electrode was then removed from the lithium foil, washed by DME, dried for 30 minutes under vacuum and studied in half lithium cell and full cell-using AC-H@S cathode. Galvanostatic cycling tests of chemically lithiated $Li_ySiO_x$-C pre-activated for 30 minutes, 1, 2 and 14 hours, as well as of a pristine $SiO_x$-C electrode, were performed in lithium half-cell using the DOL:DME 1:1 w/w, 1 mol kg$^{-1}$ LiTFSI, 1 mol kg$^{-1}$ LiNO$_3$ electrolyte at a current rate of 50 mA g$^{-1}$ in the 0.01 – 2 V voltage range. Specific capacity and current were referred to the $SiO_x$-C mass.

A further lithium-ion sulfur cell with the $Li_ySiO_x$-C/DOL:DME 1:1 w/w, 1 mol kg$^{-1}$ LiTFSI, 1 mol kg$^{-1}$ LiNO$_3$/AC-H@S configuration was assembled by using a $Li_ySiO_x$-C anode achieved by the above described chemical pre-lithiation for 48 hours, and cycled within the 0.1–2.8 V voltage window at a current rate of C/5. SEM-EDS and XRD measurements were carried out on the AC-H@S and

$SiO_x$-C electrodes employed for this cell at the pristine state, and *ex*-situ after 20 discharge/charge cycles. The SEM images were collected through a Zeiss EVO 40 microscope equipped with a $LaB_6$ thermoionic electron gun and the EDS analyses were performed by a X-ACT Cambridge Instruments analyzer. The XRD patterns were recorded through a Bruker D8 ADVANCE diffractometer employing a Cu-Kα source by performing scans between 10° and 90° in the 2θ range at a rate of 10 s step$^{-1}$ with a step size of 0.02°. Prior to perform SEM-EDS and XRD analyses the electrodes were washed with DME and dried under vacuum for 30 minutes.

**Acknowledgements**

This project/work has received funding from the European Union's Horizon 2020 research and innovation programme Graphene Flagship under grant agreement No 881603. Likewise, this research was funded by Ministerio de Economía y Competitividad (Project MAT2017-87541-R) and Junta de Andalucía (Group FQM-175). The authors also thank grant "Fondo di Ateneo per la Ricerca Locale (FAR) 2020", University of Ferrara, and the collaboration project "Accordo di Collaborazione Quadro 2015" between University of Ferrara (Department of Chemical and Pharmaceutical Sciences) and Sapienza University of Rome (Department of Chemistry).

**Institutional twitter accounts**

University of Ferrara: @UniFerrara

University of Cordoba: @Univcordoba

University of Extremadura: @infouex, @CUMe_UEx

**References**

[1]     R. Fang, S. Zhao, Z. Sun, D.-W. Wang, H.-M. Cheng, F. Li, *Adv. Mater.* **2017**, *29*, 1606823.

[2]     B. Scrosati, J. Hassoun, Y.-K. Sun, *Energy Environ. Sci.* **2011**, *4*, DOI 10.1039/c1ee01388b.

[3]     A. Manthiram, S.-H. Chung, C. Zu, *Adv. Mater.* **2015**, *27*, 1980.

[4]     D. Di Lecce, R. Verrelli, J. Hassoun, *Green Chem.* **2017**, *19*, 3442.

[5]     J. B. Goodenough, K.-S. Park, *J. Am. Chem. Soc.* **2013**, *135*, 1167.

[6]     J. Xiao, J. Z. Hu, H. Chen, M. Vijayakumar, J. Zheng, H. Pan, E. D. Walter, M. Hu, X. Deng, J. Feng, B. Y. Liaw, M. Gu, Z. D. Deng, D. Lu, S. Xu, C. Wang, J. Liu, *Nano Lett.* **2015**, *15*, 3309.

[7]     L. E. Camacho-Forero, T. W. Smith, S. Bertolini, P. B. Balbuena, *J. Phys. Chem. C* **2015**, *119*, 26828.

[8]     D. Zhu, T. Long, B. Xu, Y. Zhao, H. Hong, R. Liu, F. Meng, J. Liu, *J. Energy Chem.* **2021**, *57*, 41.

[9]     S. Wei, S. Inoue, D. Di Lecce, Z. Li, Y. Tominaga, J. Hassoun, *ChemElectroChem* **2020**, *7*, 2344.

[10]    V. Marangon, Y. Tominaga, J. Hassoun, *J. Power Sources* **2020**, *449*, 227508.

[11]    S. Xiong, K. Xie, Y. Diao, X. Hong, *J. Power Sources* **2014**, *246*, 840.

[12]    A. Rosenman, R. Elazari, G. Salitra, E. Markevich, D. Aurbach, A. Garsuch, *J. Electrochem. Soc.* **2015**, *162*, A470.

[13]    W. Li, Z. Liang, Z. Lu, H. Yao, Z. W. Seh, K. Yan, G. Zheng, Y. Cui, *Adv. Energy Mater.* **2015**, *5*, DOI 10.1002/aenm.201500211.

[14]    L. Carbone, T. Coneglian, M. Gobet, S. Munoz, M. Devany, S. Greenbaum, J. Hassoun, *J.*

*Power Sources* **2018**, *377*, 26.

[15] J. Hou, X. Tu, X. Wu, M. Shen, X. Wang, C. Wang, C. Cao, H. Pang, G. Wang, *Chem. Eng. J.* **2020**, *401*, 126141.

[16] G. Zhou, Y. Zhao, A. Manthiram, *Adv. Energy Mater.* **2015**, *5*, 1402263.

[17] L. Guan, H. Hu, L. Li, Y. Pan, Y. Zhu, Q. Li, H. Guo, K. Wang, Y. Huang, M. Zhang, Y. Yan, Z. Li, X. Teng, J. Yang, J. Xiao, Y. Zhang, X. Wang, M. Wu, *ACS Nano* **2020**, *14*, 6222.

[18] D. Di Lecce, V. Marangon, A. Benítez, Á. Caballero, J. Morales, E. Rodríguez-Castellón, J. Hassoun, *J. Power Sources* **2019**, *412*, DOI 10.1016/j.jpowsour.2018.11.068.

[19] L. Wang, A. Abraham, D. M. Lutz, C. D. Quilty, E. S. Takeuchi, K. J. Takeuchi, A. C. Marschilok, *ACS Sustain. Chem. Eng.* **2019**, *7*, 5209.

[20] L. Carbone, M. Gobet, J. Peng, M. Devany, B. Scrosati, S. Greenbaum, J. Hassoun, *ACS Appl. Mater. Interfaces* **2015**, *7*, 13859.

[21] J. Wang, S. Y. Chew, Z. W. Zhao, S. Ashraf, D. Wexler, J. Chen, S. H. Ng, S. L. Chou, H. K. Liu, *Carbon N. Y.* **2008**, *46*, 229.

[22] L. Wang, J. Liu, S. Yuan, Y. Wang, Y. Xia, *Energy Environ. Sci.* **2016**, *9*, 224.

[23] J. Hassoun, B. Scrosati, *Adv. Mater.* **2010**, *22*, DOI 10.1002/adma.201002584.

[24] L. Carbone, S. G. Greenbaum, J. Hassoun, **2017**, *1*, 228.

[25] W. Xu, J. Wang, F. Ding, X. Chen, E. Nasybulin, Y. Zhang, J.-G. Zhang, *Energy Environ. Sci.* **2014**, *7*, 513.

[26] J.-L. Brédas, J. M. Buriak, F. Caruso, K.-S. Choi, B. A. Korgel, M. R. Palacín, K. Persson, E. Reichmanis, F. Schüth, R. Seshadri, M. D. Ward, *Chem. Mater.* **2019**, *31*, 8577.

[27] *Angew. Chemie Int. Ed.* **2019**, *58*, 16723.

[28] D. Di Lecce, P. Andreotti, M. Boni, G. Gasparro, G. Rizzati, J.-Y. Hwang, Y.-K. Sun, J. Hassoun, *ACS Sustain. Chem. Eng.* **2018**, *6*, 3225.

[29] J.-M. Kim, J.-K. Hwang, Y.-K. Sun, J. Hassoun, *Sustain. Energy Fuels* **2019**, *3*, DOI 10.1039/c9se00259f.

[30] J. Hassoun, P. Ochal, S. Panero, G. Mulas, C. Bonatto Minella, B. Scrosati, *J. Power Sources* **2008**, *180*, DOI 10.1016/j.jpowsour.2008.01.059.

[31] G. Wang, F. Li, D. Liu, D. Zheng, C. J. Abeggien, Y. Luo, X.-Q. Yang, T. Ding, D. Qu, *Energy Storage Mater.* **2020**, *24*, 147.

[32] A. Benítez, D. Di Lecce, G. A. Elia, Á. Caballero, J. Morales, J. Hassoun, *ChemSusChem* **2018**, *11*, 1512.

[33] S. Lin, Y. Yan, Z. Cai, L. Liu, X. Hu, *Small* **2018**, *14*, 1800616.

[34] S. Thieme, J. Brückner, A. Meier, I. Bauer, K. Gruber, J. Kaspar, A. Helmer, H. Althues, M. Schmuck, S. Kaskel, *J. Mater. Chem. A* **2015**, *3*, 3808.

[35] Y. Yao, H. Wang, H. Yang, S. Zeng, R. Xu, F. Liu, P. Shi, Y. Feng, K. Wang, W. Yang, X. Wu, W. Luo, Y. Yu, *Adv. Mater.* **2020**, *32*, 1905658.

[36] J. Jiang, Q. Fan, Z. Zheng, M. R. Kaiser, Q. Gu, S. Chou, K. Konstantinov, J. Wang, *ACS Appl. Energy Mater.* **2020**, *3*, 6447.

[37] L. Liu, F. Xie, J. Lyu, T. Zhao, T. Li, B. G. Choi, *J. Power Sources* **2016**, *321*, 11.

[38] G. A. Elia, F. Nobili, R. Tossici, R. Marassi, A. Savoini, S. Panero, J. Hassoun, *J. Power Sources* **2015**, *275*, DOI 10.1016/j.jpowsour.2014.10.144.

[39] M. Ge, C. Cao, G. M. Biesold, C. D. Sewell, S. Hao, J. Huang, W. Zhang, Y. Lai, Z. Lin,

*Adv. Mater.* **2021**, *33*, 2004577.

[40] K. Zhao, L. Zhang, R. Xia, Y. Dong, W. Xu, C. Niu, L. He, M. Yan, L. Qu, L. Mai, *Small* **2016**, *12*, 588.

[41] H. Woo, S. Wi, J. Kim, J. Kim, S. Lee, T. Hwang, J. Kang, J. Kim, K. Park, B. Gil, S. Nam, B. Park, *Carbon N. Y.* **2018**, *129*, 342.

[42] H.-W. Yang, D. I. Lee, N. Kang, J.-K. Yoo, S.-T. Myung, J. Kim, S.-J. Kim, *J. Power Sources* **2018**, *400*, 613.

[43] J. Ge, Q. Tang, H. Shen, F. Zhou, H. Zhou, W. Yang, J. Hong, B. Xu, J. Saddique, *Appl. Surf. Sci.* **2021**, *552*, 149446.

[44] G. Carbonari, F. Maroni, S. Gabrielli, A. Staffolani, R. Tossici, A. Palmieri, F. Nobili, *ChemElectroChem* **2019**, *6*, 1915.

[45] Y. Wu, S. Zhang, J. Hao, H. Liu, X. Wu, J. Hu, M. P. Walsh, T. J. Wallington, K. M. Zhang, S. Stevanovic, *Sci. Total Environ.* **2017**, *574*, 332.

[46] J. Chen, C. Li, Z. Ristovski, A. Milic, Y. Gu, M. S. Islam, S. Wang, J. Hao, H. Zhang, C. He, H. Guo, H. Fu, B. Miljevic, L. Morawska, P. Thai, Y. F. LAM, G. Pereira, A. Ding, X. Huang, U. C. Dumka, *Sci. Total Environ.* **2017**, *579*, 1000.

[47] J. Wang, P. Nie, B. Ding, S. Dong, X. Hao, H. Dou, X. Zhang, *J. Mater. Chem. A* **2017**, *5*, 2411.

[48] S. Dühnen, J. Betz, M. Kolek, R. Schmuch, M. Winter, T. Placke, *Small Methods* **2020**, *4*, 2000039.

[49] F. Wang, D. Ouyang, Z. Zhou, S. J. Page, D. Liu, X. Zhao, *J. Energy Chem.* **2021**, *57*, 247.

[50] H. Yao, G. Zheng, W. Li, M. T. McDowell, Z. Seh, N. Liu, Z. Lu, Y. Cui, *Nano Lett.* **2013**, *13*, 3385.

[51] J. Zhou, Y. Guo, C. Liang, J. Yang, J. Wang, Y. Nuli, *Electrochim. Acta* **2018**, *273*, 127.

[52] N. Moreno, A. Caballero, L. Hernán, J. Morales, *Carbon N. Y.* **2014**, *70*, 241.

[53] M. K. Rybarczyk, H.-J. Peng, C. Tang, M. Lieder, Q. Zhang, M.-M. Titirici, *Green Chem.* **2016**, *18*, 5169.

[54] C. Hernández-Rentero, R. Córdoba, N. Moreno, A. Caballero, J. Morales, M. Olivares-Marín, V. Gómez-Serrano, *Nano Res.* **2018**, *11*, 89.

[55] Y. Yan, M. Shi, Y. Wei, C. Zhao, M. Carnie, R. Yang, Y. Xu, *J. Alloys Compd.* **2018**, *738*, 16.

[56] A. Y. Tesio, J. L. Gómez-Cámer, J. Morales, A. Caballero, *ChemSusChem* **2020**, *13*, 3439.

[57] J. Han, X. Chen, B. Xi, H. Mao, J. Feng, S. Xiong, *ChemistrySelect* **2018**, *3*, 10175.

[58] L. Borchardt, Steve Buscaglia, Daniela Barbero Vignola, Giulia Maroni, Michele Marelli, *JRC Sci. policy Rep.* **2020**, *EUR 30452*, DOI 10.2760/030575.

[59] G. A. Elia, J. Hassoun, *ChemElectroChem* **2017**, *4*, 2164.

[60] C. Hernández-Rentero, V. Marangon, M. Olivares-Marín, V. Gómez-Serrano, Á. Caballero, J. Morales, J. Hassoun, *J. Colloid Interface Sci.* **2020**, *573*, 396.

[61] P. Lian, X. Zhu, S. Liang, Z. Li, W. Yang, H. Wang, *Electrochim. Acta* **2010**, *55*, 3909.

[62] A. C. Ferrari, J. C. Meyer, V. Scardaci, C. Casiraghi, M. Lazzeri, F. Mauri, S. Piscanec, D. Jiang, K. S. Novoselov, S. Roth, A. K. Geim, *Phys. Rev. Lett.* **2006**, *97*, 187401.

[63] A. Benítez, D. Di Lecce, Á. Caballero, J. Morales, E. Rodríguez-Castellón, J. Hassoun, *J. Power Sources* **2018**, *397*, 102.

[64] T. Xing, L. H. Li, L. Hou, X. Hu, S. Zhou, R. Peter, M. Petravic, Y. Chen, *Carbon N. Y.* **2013**, *57*, 515.

[65] G. Li, J. Sun, W. Hou, S. Jiang, Y. Huang, J. Geng, *Nat. Commun.* **2016**, *7*, 10601.

[66] V. Marangon, J. Hassoun, *Energy Technol.* **2019**, *7*, 1900081.

[67] W. G. Wang, X. Wang, L. Y. Tian, Y. L. Wang, S. H. Ye, *J. Mater. Chem. A* **2014**, *2*, 4316.

[68] M. Chen, S. Jiang, S. Cai, X. Wang, K. Xiang, Z. Ma, P. Song, A. C. Fisher, *Chem. Eng. J.* **2017**, *313*, 404.

[69] V. Marangon, D. Di Lecce, F. Orsatti, D. J. L. Brett, P. R. Shearing, J. Hassoun, *Sustain. Energy Fuels* **2020**, *4*, 2907.

[70] M. Olivares-Marín, C. Fernández-González, A. Macías-García, V. Gómez-Serrano, *Energy & Fuels* **2007**, *21*, 2942.

[71] D. Di Lecce, V. Marangon, W. Du, D. J. L. Brett, P. R. Shearing, J. Hassoun, *J. Power Sources* **2020**, *472*, 228424.

[72] A. Benítez, V. Marangon, C. Hernández-Rentero, Á. Caballero, J. Morales, J. Hassoun, *Mater. Chem. Phys.* **2020**, *255*, DOI 10.1016/j.matchemphys.2020.123484.

[73] B. BOUKAMP, *Solid State Ionics* **1986**, *18–19*, 136.

[74] B. BOUKAMP, *Solid State Ionics* **1986**, *20*, 31.

[75] C. Shen, M. Ge, A. Zhang, X. Fang, Y. Liu, J. Rong, C. Zhou, *Nano Energy* **2016**, *19*, 68.

[76] M. Agostini, J. Hassoun, J. Liu, M. Jeong, H. Nara, T. Momma, T. Osaka, Y.-K. Sun, B. Scrosati, *ACS Appl. Mater. Interfaces* **2014**, *6*, 10924.

[77] J. Hassoun, D. J. Lee, Y. K. Sun, B. Scrosati, *Solid State Ionics* **2011**, *202*, 36.

[78] J. Hassoun, K.-S. Lee, Y.-K. Sun, B. Scrosati, *J. Am. Chem. Soc.* **2011**, *133*, 3139.

[79] Q. Liu, A. Cresce, M. Schroeder, K. Xu, D. Mu, B. Wu, L. Shi, F. Wu, *Energy Storage Mater.* **2019**, *17*, 366.

[80] V. Marangon, C. Hernandez-Rentero, S. Levchenko, G. Bianchini, D. Spagnolo, A. Caballero, J. Morales, J. Hassoun, S. Cathode, *ACS Appl. Energy Mater.* **2020**, *3*, 12263.

[81] M. R. Busche, P. Adelhelm, H. Sommer, H. Schneider, K. Leitner, J. Janek, *J. Power Sources* **2014**, *259*, 289.

[82] A. Benítez, Á. Caballero, E. Rodríguez-Castellón, J. Morales, J. Hassoun, *ChemistrySelect* **2018**, *3*, 10371.

[83] A. Benítez, F. Luna-Lama, A. Caballero, E. Rodríguez-Castellón, J. Morales, *J. Energy Chem.* **2021**, *62*, 295.

**List of Tables**

**Table 1.** NLLS analyses carried out on the EIS Nyquist Plots reported in Figure 3b recorded upon CV test performed on a Li/DOL:DME 1:1 w/w, 1 mol kg$^{-1}$ LiTFSI, 1 mol kg$^{-1}$ LiNO$_3$/AC-H@S cell. The analyses were carried out by applying the NLLS method through a Boukamp tool.[73,74]

**List of Figures**

**Figure 1. (a)** X-ray diffraction (XRD) pattern, **(b)** Raman spectrum, and **(c)** thermogravimetric analysis (TGA) with corresponding derivative thermogravimetric (DTG) curve (bottom panel, orange left *y*-axis) of the AC-H@S composite powder. XRD reference data for elemental sulfur (S$_8$, PDF # 8-247, orange) and graphite (PDF # 41-1487, black) are also reported for comparison. TGA was carried out under N$_2$ atmosphere in the 30 – 800 °C temperature range at 10 °C min$^{-1}$. **(d-i)** Morphological analyses of AC-H@S powder samples. In detail: **(d)** Transmission electron microscopy (TEM) and **(e)** scanning electron microscopy (SEM) images; **(f)** additional TEM image and **(g-i)** corresponding energy dispersive X-ray spectroscopy (EDS) elemental maps for **(g)** carbon, **(h)** sulfur and **(i)** phosphorus.

**Figure 2. (a)** Cyclic voltammetry (CV) and **(b)** electrochemical impedance spectroscopy (EIS) measurements performed on a Li/DOL:DME 1:1 w/w, 1 mol kg$^{-1}$ LiTFSI, 1 mol kg$^{-1}$ LiNO$_3$/AC-H@S cell. CV potential range: 1.8 – 2.8 V *vs.* Li$^+$/Li, scan rate: 0.05 mV s$^{-1}$. Impedance spectra were recorded in the 500 kHz – 100 mHz frequency range (signal amplitude: 10 mV) at the open circuit voltage (OCV, inset in panel b) of the cell and after 1, 5 and 10 CV cycles.

**Figure 3. (a-d)** Galvanostatic tests performed on Li/DOL:DME 1:1 w/w, 1 mol kg$^{-1}$ LiTFSI, 1 mol kg$^{-1}$ LiNO$_3$/AC-H@S half-cells. In detail: **(a,c)** selected voltage profiles and **(b,d)** corresponding cycling trend (right *y*-axis in panel **(d)** refers to coulombic efficiency) related to tests carried out **(a,b)** at increasing currents employing the C/10, C/8, C/5, C/3, C/2, 1C and 2C rates and **(c,d)** at the constant current rate of C/3 (1C = 1675 mA g$_S^{-1}$). Voltage ranges: 1.9 – 2.8 V from C/10 to C/2 and

1.8 – 2.8 V for 1C and 2C rates. **(e-f)** Galvanostatic tests performed on $Li_ySiO_x$-C/DOL:DME 1:1 w/w, 1 mol kg$^{-1}$ LiTFSI, 1 mol kg$^{-1}$ LiNO$_3$/AC-H@S full-cell at a C/5 current rate (1C = 1675 mA gs$^{-1}$). In detail: **(e)** selected voltage profiles and **(f)** corresponding cycling trend with coulombic efficiency in right *y*-axis. Voltage range: 0.1 – 2.8 V. The N/P ratio between the $Li_ySiO_x$-C and AC-H@S electrodes was tuned to a value of 1.04. The anode was electrochemically pre-lithiated at 50 mA g$^{-1}$ for over 30 cycles in the 0.01 - 2.0 V voltage range.

**Figure 4. (a-f)** Voltage profiles related to the first cycle of **(a)** a Li/DOL:DME 1:1 w/w, 1 mol kg$^{-1}$ LiTFSI, 1 mol kg$^{-1}$ LiNO$_3$/SiO$_x$-C pristine cell and **(b-e)** Li/DOL:DME 1:1 w/w, 1 mol kg$^{-1}$ LiTFSI, 1 mol kg$^{-1}$ LiNO$_3$/Li$_y$SiO$_x$-C cells employing Li$_y$SiO$_x$-C electrodes chemically pre-lithiated (activated) at various times, that is, **(b)** 30 minutes, **(c)** 1 hour, **(d)** 2 hours and **(e)** 14 hours (see experimental section for details). Panel **(f)** displays a histogram representation of the discharge capacity values obtained from the cycling tests reported in panels **(a-e)**. All the cycling measurements were carried out at a current rate of 50 mA g$^{-1}$ (referred to the pristine SiO$_x$-C mass) in the 0.01 – 2 V voltage range. **(g)** Voltage profiles related to the full cell exploiting the Li$_y$SiO$_x$-C/DOL:DME 1:1 w/w, 1 mol kg$^{-1}$ LiTFSI, 1 mol kg$^{-1}$ LiNO$_3$/AC-H@S configuration galvanostatically cycled at the constant rate of C/5 (1C = 1675 mA gs$^{-1}$) in the 0.1 – 2.8 V voltage range. The anode was chemically pre-lithiated by employing an activation time of 48 hours (see experimental section for details).

**Figure 5. (a-d)** SEM images recorded on **(a,c)** AC-H@S and **(b,d)** SiO$_x$-C electrodes at **(a,b)** the prsitine state and **(c,d)** after 20 discharge/charge cycles in a Li$_y$SiO$_x$-C/DOL:DME 1:1 w/w, 1 mol kg$^{-1}$ LiTFSI, 1 mol kg$^{-1}$ LiNO$_3$/AC-H@S cell (see the corresponding voltage profiles in Fig. 4g) cycled at a constant current rate of C/5 (1C = 1675 mA gs$^{-1}$) in the 0.1 – 2.8 V voltage range; XRD patterns of the **(e)** AC-H@S and **(f)** SiO$_x$-C electrodes before and after cycling. XRD reference data for elemental sulfur (S$_8$, panel **(e)**, PDF # 8-247) and copper (Cu, panel **(f)**, PDF # 4-836) are also reported for comparison.

| Cell condition | Equivalent circuit | $R_1$ [Ω] | $R_2$ [Ω] | $R_3$ [Ω] | $R_{tot}$ $\sum_{i=1}^{3} R_i$ [Ω] | $\chi^2$ |
|---|---|---|---|---|---|---|
| OCV | $R_e(R_1Q_1)(R_2Q_2)Q_w$ | 32 ± 4 | 8.0 ± 3.8 | / | 40 ± 8 | 1×10$^{-4}$ |
| 1 CV cycle | $R_e(R_1Q_1)(R_2Q_2)Q_w$ | 6.1 ± 0.1 | 2.3 ± 0.1 | / | 8.4 ± 0.2 | 6×10$^{-6}$ |
| 5 CV cycles | $R_e(R_1Q_1)(R_2Q_2)(R_3Q_3)Q_w$ | 0.8 ± 0.3 | 3.3 ± 0.3 | 1.5 ± 0.3 | 5.6 ± 0.9 | 8×10$^{-5}$ |
| 10 CV cycles | $R_e(R_1Q_1)(R_2Q_2)(R_3Q_3)Q_w$ | 1.0 ± 0.1 | 3.8 ± 0.2 | 1.3 ± 0.1 | 6.1 ± 0.4 | 1×10$^{-5}$ |

**Table 1**

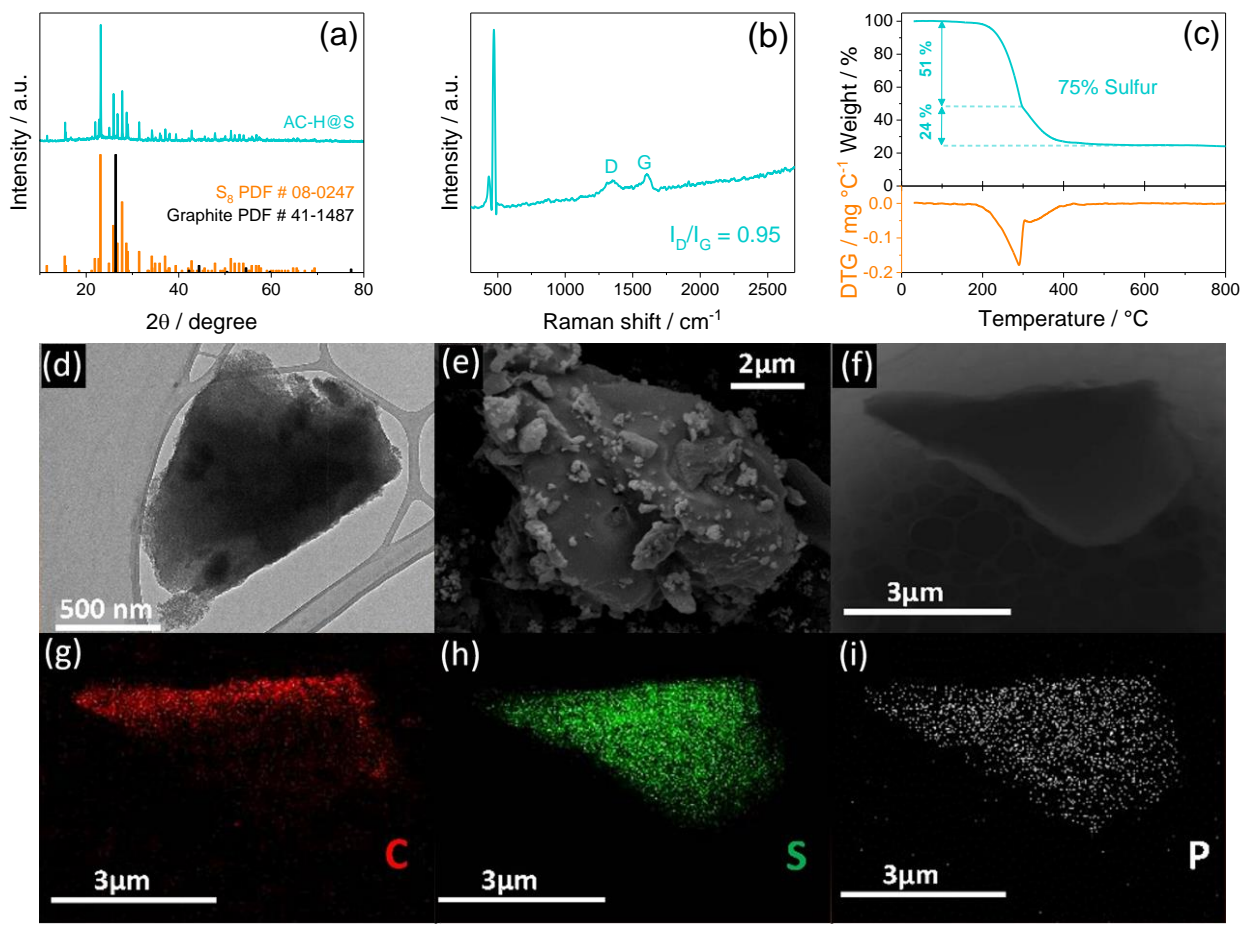

**Figure 1**

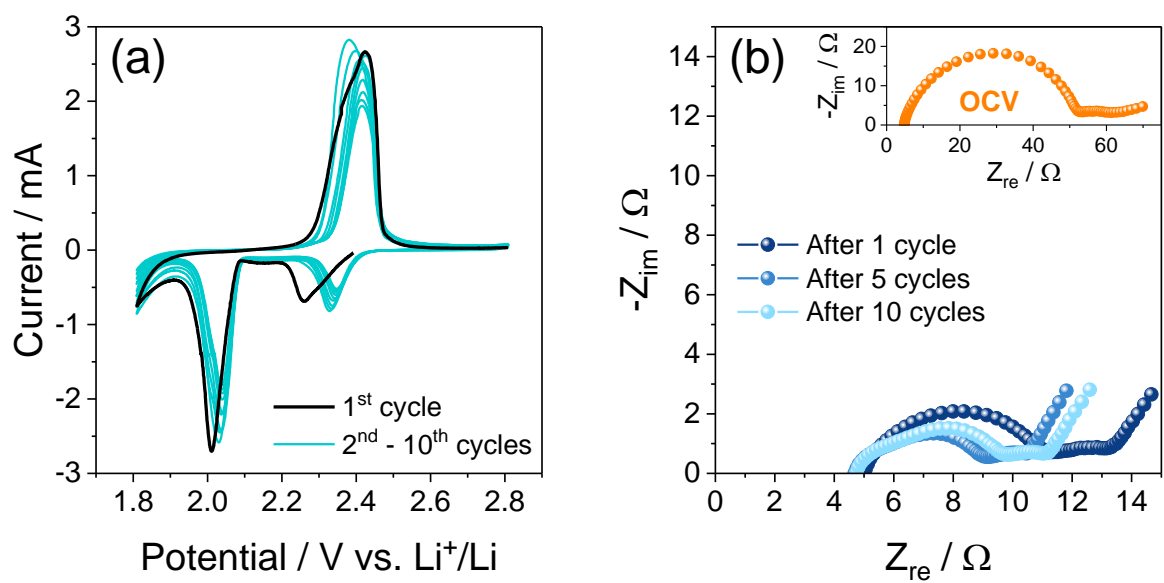

**Figure 2**

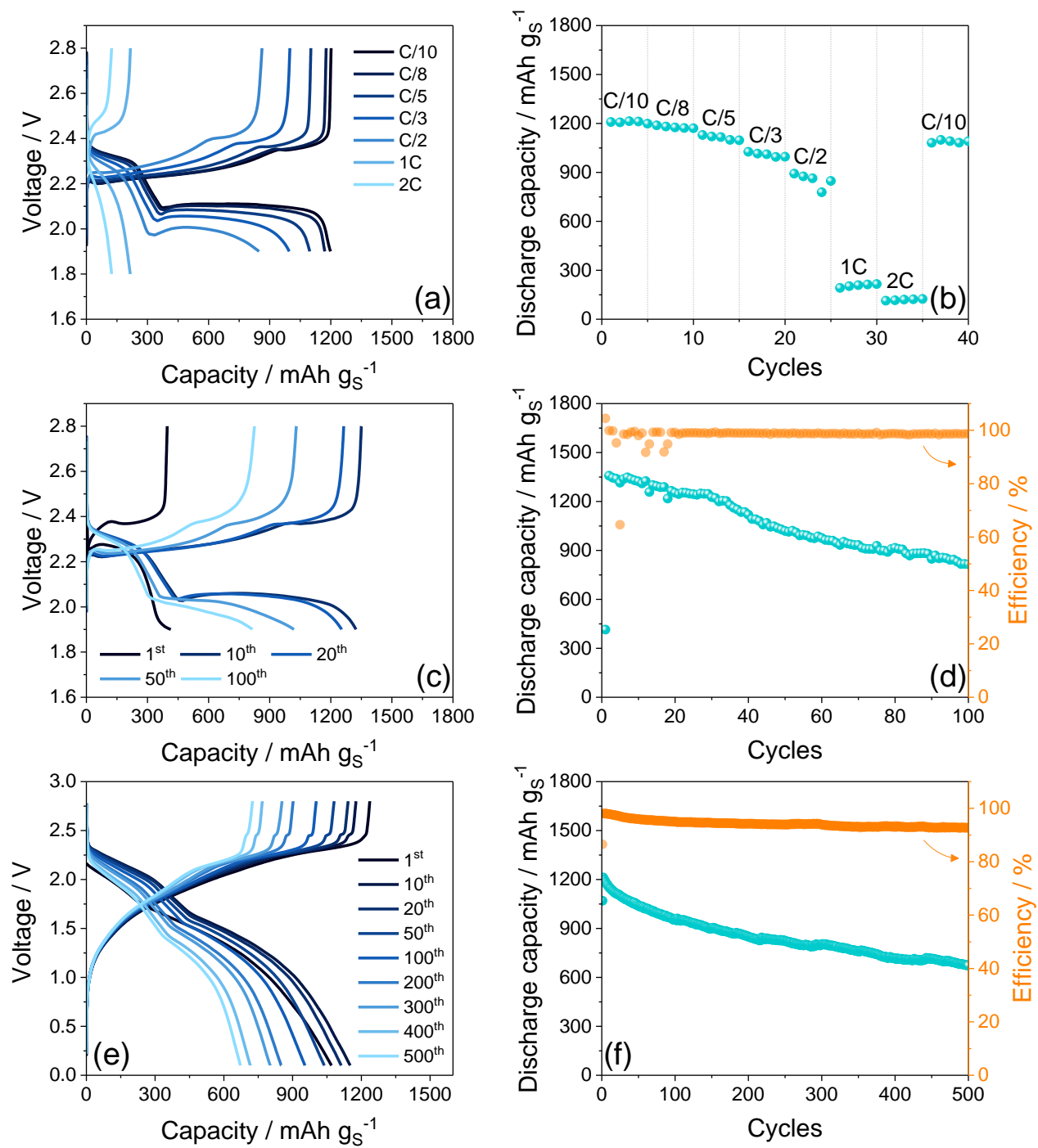

**Figure 3**

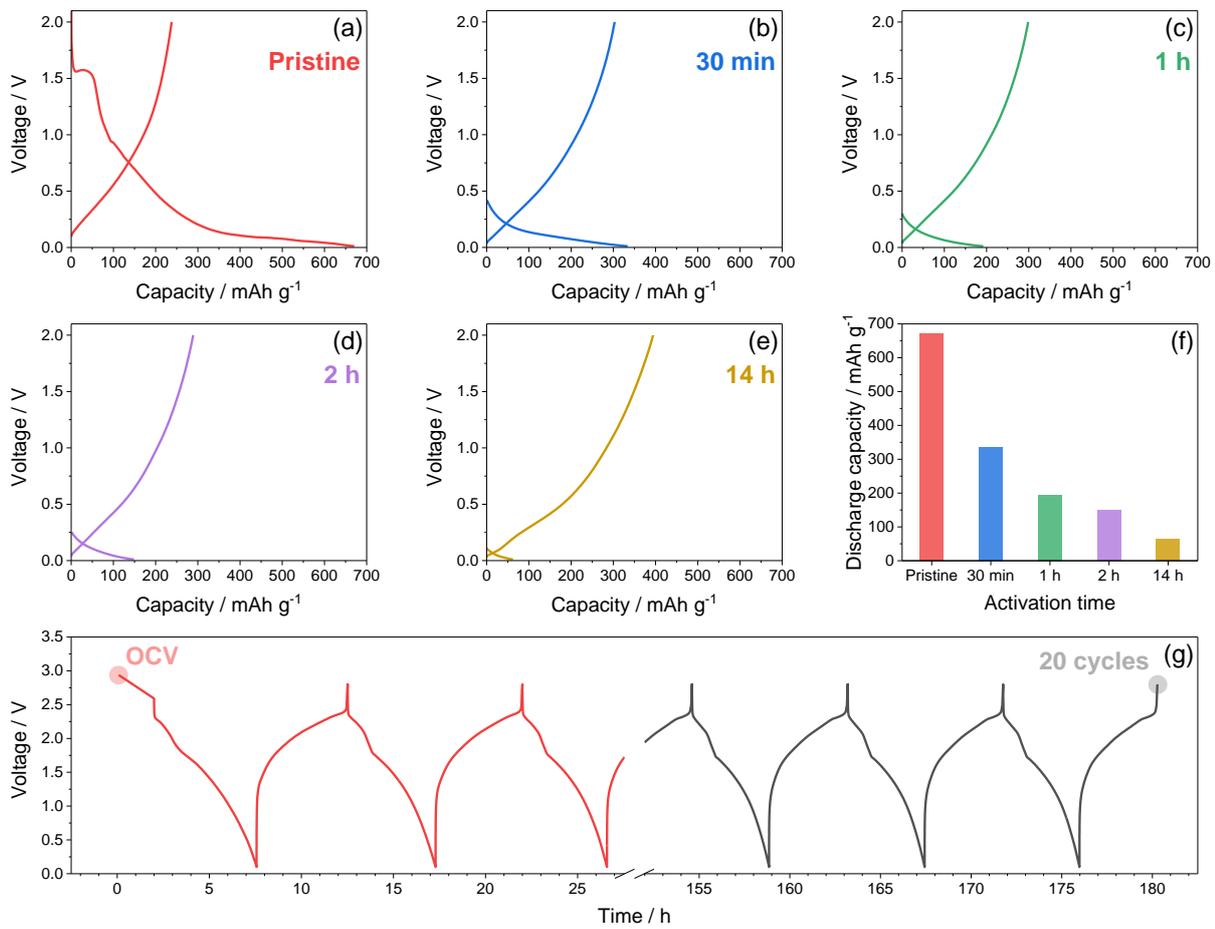

**Figure 4**

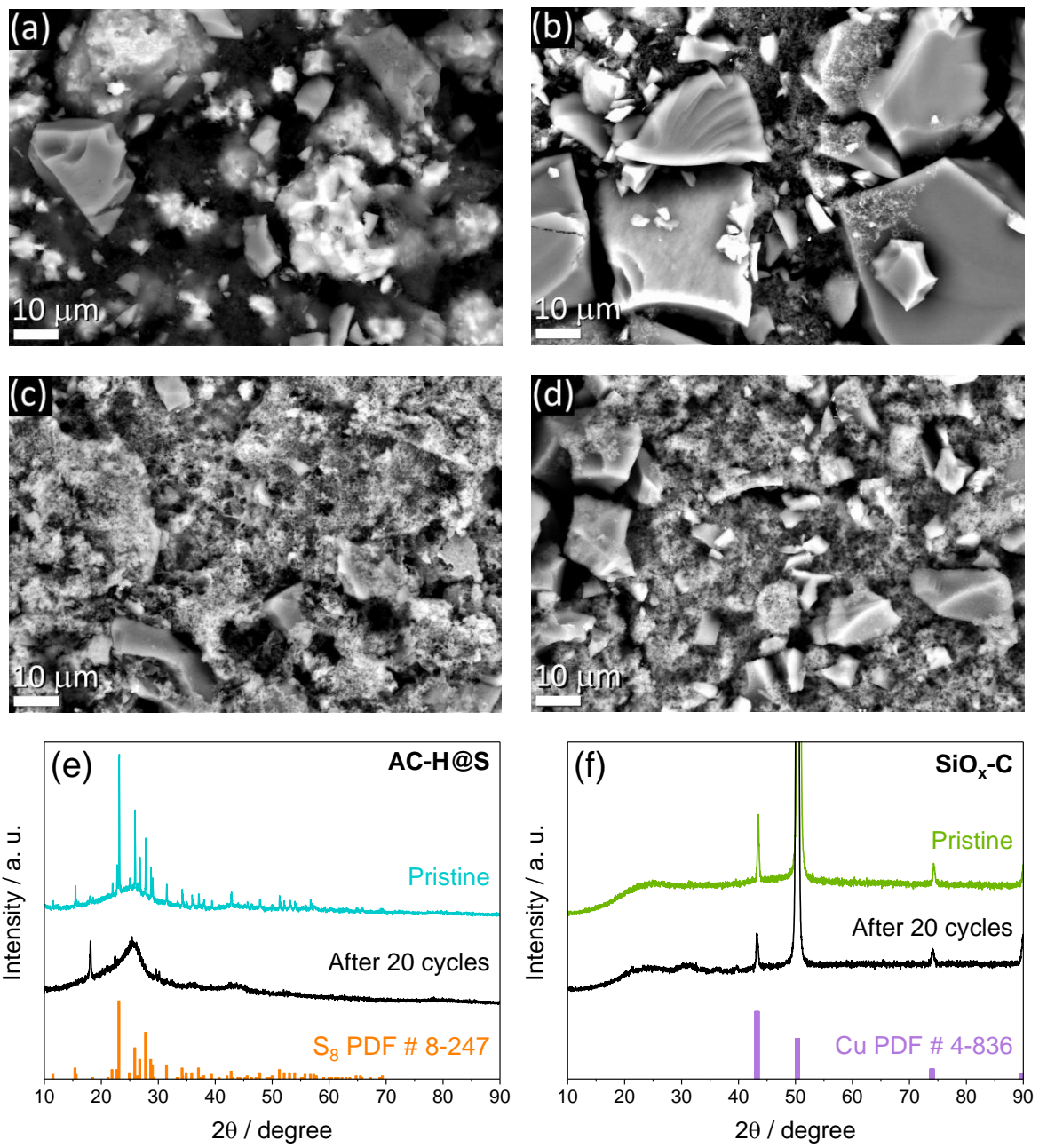

**Figure 5**

**A Stable High Capacity Lithium-Ion Battery Using a Biomass-Derived Sulfur-Carbon Cathode and Lithiated Silicon Anode**

Vittorio Marangon[a,†], Celia Hernández-Rentero[b,†], Dr. Mara Olivares-Marín[c], Prof. Vicente Gómez-Serrano[d], Prof. Álvaro Caballero[b], Prof. Julián Morales[b]*, Prof. Jusef Hassoun[a,e,f]*

[a] Department of Chemical, Pharmaceutical and Agricultural Sciences, University of Ferrara, Via Fossato di Mortara 17, Ferrara 44121, Italy

[b] Department of Química Inorgánica e Ingeniería Química, Instituto de Química Fina y Nanoquímica, University of Córdoba, 14071 Córdoba, Spain

[c] Department of Ingeniería Mecánica, Energética y de los Materiales, University of Extremadura, Centro Universitario de Mérida, 06800 Mérida, Spain

[d] Department of Química Inorgánica, Facultad de Ciencias, University of Extremadura, 06006 Badajoz, Spain

[e] Graphene Labs, Istituto Italiano di Tecnologia, Via Morego 30 – 16163 Genova, Italy.

[f] National Interuniversity Consortium of Materials Science and Technology (INSTM), University of Ferrara Research Unit, University of Ferrara, Via Fossato di Mortara, 17, 44121, Ferrara, Italy.

[†] Authors equally contributed.

*Corresponding Authors. E-mail addresses: iq1mopaj@uco.es (Julián Morales), jusef.hassoun@unife.it, jusef.hassoun@iit.it (Jusef Hassoun).

**Supporting Information**

The active material loadings of about 5.3 mg cm$^{-2}$ for SiO$_x$-C and 1.3 mg cm$^{-2}$ for AC-H@S allow the achievement of a N/P ratio nearby 1 as shown in Figure S1 which reports steady state voltage profiles of the two materials in lithium half-cell (panel a and b, respectively). It is worth mentioning that the different masses lead to almost the same surface geometrical capacity (that is, of about 1.7 mAh cm$^{-2}$) due to the different specific capacity of the active materials.

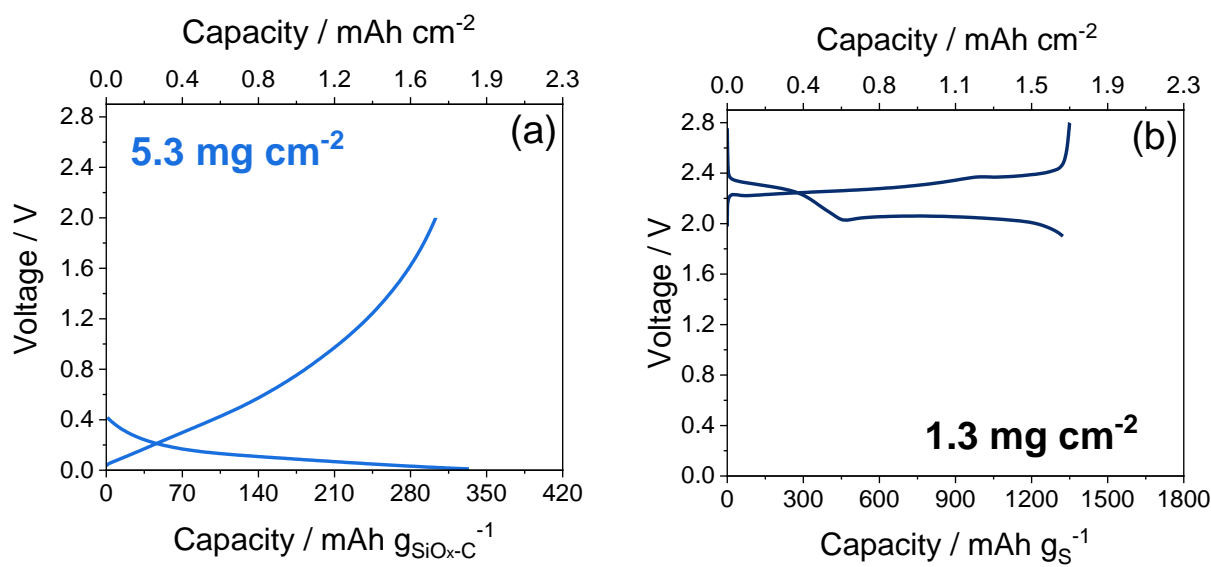

**Figure S1.** Steady state voltage profile of **(a)** SiO$_x$-C and **(b)** AC-H@S in lithium half-cell reported to determine the N/P ratio of the full-cell assembled using the two materials and reported in the manuscript (see discussion of Figure 3 and 4). Current values and voltage limits: 50 mA g$^{-1}$, 0.01 – 2 V for Li/DOL:DME 1:1 w/w, 1 mol kg$^{-1}$ LiTFSI, 1 mol kg$^{-1}$ LiNO$_3$/SiO$_x$-C cell; and 170 mA g$^{-1}$, 1.8 – 2.8 V for Li/DOL:DME 1:1 w/w, 1 mol kg$^{-1}$ LiTFSI, 1 mol kg$^{-1}$ LiNO$_3$/AC-H@S cell

A $SiO_x$-C electrode chemically pre-activated (lithiated) for 48 hours (see experimental section) was employed to assemble a $Li_ySiO_x$-C/AC-H@S full-cell which performed 20 discharge/charge cycles at the constant rate of C/5 (1C = 1675 mA $g_S^{-1}$), as displayed in Figure 4 of the manuscript. Upon cycling, the electrodes were recovered from the disassembled cell to perform SEM-EDS and XRD analyses on their surfaces. Figure 5 in the Manuscript reports the comparison of SEM images acquired on the surfaces of pristine and cycled AC-H@S and $SiO_x$-C electrodes, while the corresponding EDS elemental maps are shown in Figure S2 for AC-H@S and Figure S3 for $SiO_x$-C. The EDS elemental maps of the pristine AC-H@S electrode reveal sulfur aggregates (Fig. S2b) and a relevant presence of phosphorous residues due the activation of the carbon by $H_3PO_4$ (Fig. S3c). The C (Fig. S2a) and F (Fig. S2d) maps display the uniform distribution of the carbon and the PVDF polymer binder, respectively, used in the electrode slurry (see experimental section). It is worth mentioning that the carbonaceous polymeric chain of the PVDF may also contribute to the C signals in Fig. S2a. After cycling, heterogeneously phosphorous-covered particles can be observed (Fig. S2g), while carbon (Fig. S2e), sulfur (Fig. S2f), fluorine (Fig. S2h) and oxygen (inset in Fig. S2h) are uniformly distributed across the electrode surface to form a layer. This layer, which contributes to the SEI film formation upon cycling, is due to electrodeposition of amorphous sulfur upon charge and partial decomposition of the DOL:DME-LiTFSI electrolyte solution.[1] The pristine $SiO_x$-C electrode exhibits defined domains of the carbon matrices entrapping $SiO_x$ particles, as evidenced by EDS elemental maps of C (Fig. S3a), Si (Fig. S3b) and O (Fig. S3c),[2] the sizes of which decrease upon cycling in full-cell (Fig. S3f) concomitantly to the formation of a uniform layer composed by carbon (Fig. S3e), oxygen (Fig. S3g), fluorine (Fig. S3h) and sulfur (inset Fig. S3h). The contribute of fluorine can also be observed in the pristine $SiO_x$-C electrode (Fig. S3d) due to the above mentioned PVDF (see experimental section).

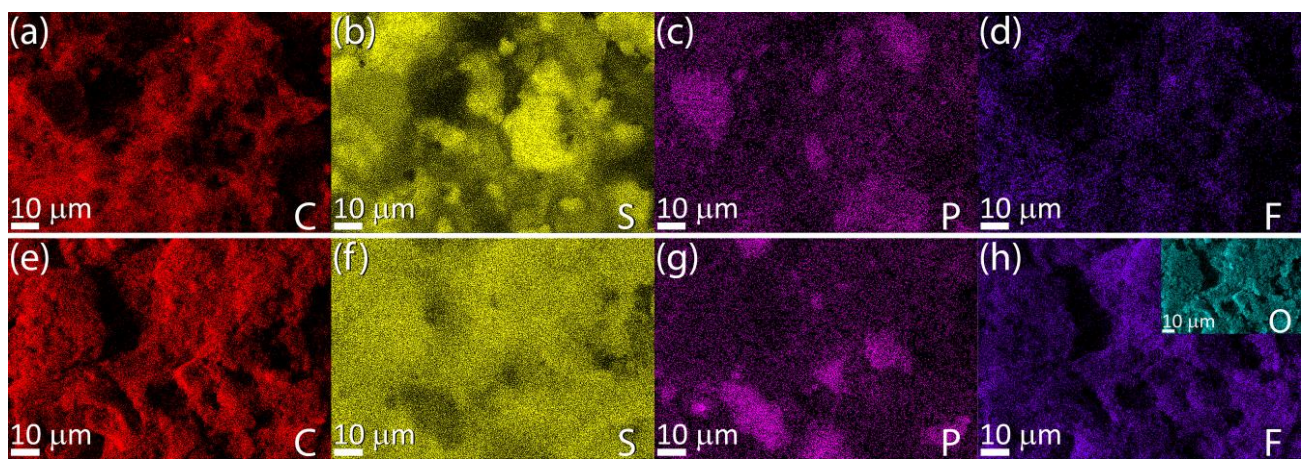

**Figure S2.** EDS elemental maps recorded on AC-H@S electrodes surface either **(a-d)** at the pristine state or **(e-h)** upon 20 cycles in a Li$_y$SiO$_x$-C/DOL:DME 1:1 w/w, 1 mol kg$^{-1}$ LiTFSI, 1 mol kg$^{-1}$ LiNO$_3$/AC-H@S full-cell at the constant current rate of C/5 (1C = 1675 mA g$_S^{-1}$) in the 0.1 – 2.8 V voltage range (see Fig. 4g in the manuscript for the corresponding voltage profiles).

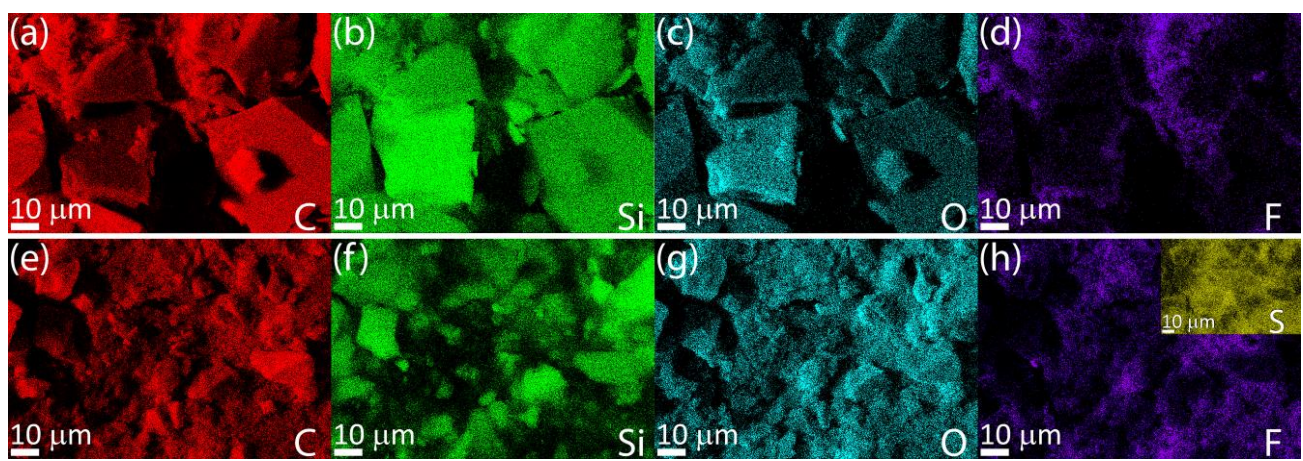

**Figure S3.** EDS elemental maps recorded on SiO$_x$-C electrodes surface either **(a-d)** at the pristine state or **(e-h)** upon 20 cycles in a Li$_y$SiO$_x$-C/DOL:DME 1:1 w/w, 1 mol kg$^{-1}$ LiTFSI, 1 mol kg$^{-1}$ LiNO$_3$/AC-H@S full-cell at a constant current rate of C/5 (1C = 1675 mA g$_S^{-1}$) in the 0.1 – 2.8 V voltage range (see Fig. 4g in the manuscript for the corresponding voltage profiles).

**References**

[1] Q. Liu, A. Cresce, M. Schroeder, K. Xu, D. Mu, B. Wu, L. Shi, F. Wu, *Energy Storage Mater.* **2019**, *17*, 366.

[2] G. A. Elia, J. Hassoun, *ChemElectroChem* **2017**, *4*, 2164.